\documentclass[repr,laps,pra,twocolumn,floatfix,showpacs]{revtex4}
\usepackage[T1]{fontenc}
\usepackage[american]{babel}
\usepackage[colorlinks=true,citecolor=blue,linkcolor=blue,urlcolor=blue]{hyperref}
\usepackage{color}
\usepackage[dvips]{graphicx}
\usepackage{amsmath,amssymb,amsthm,amsfonts}

\def\be{\begin{equation}}
\def\ee{\end{equation}}
\def\ba{\begin{eqnarray}}
\def\ea{\end{eqnarray}}

\begin{document}
\title{Comment on ``Geometric derivation of the quantum speed limit''}
\author{Marcin Zwierz} \email{zwierz.marcin@gmail.com}
\affiliation{Centre for Quantum Computation and Communication Technology (Australian Research Council), Centre for Quantum Dynamics, Griffith University, Brisbane, QLD 4111, Australia}
\date{\today}
  
\begin{abstract} 
Recently, Jones and Kok [P. J. Jones and P. Kok, Phys. Rev. A \textbf{82}, 022107 (2010)] presented alternative geometric derivations of the Mandelstam-Tamm [L. Mandelstam and I. Tamm, J. Phys. (USSR) \textbf{9}, 249 (1945)] and Margolus-Levitin [N. Margolus and L. B. Levitin, Phys. D \textbf{120}, 188 (1998)] inequalities for the quantum speed of dynamical evolution. The Margolus-Levitin inequality followed from an upper bound on the rate of change of the statistical distance between two arbitrary pure quantum states. We show that the derivation of this bound is incorrect. Subsequently, we provide two upper bounds on the rate of change of the statistical distance, expressed in terms of the standard deviation of the generator $K$ and its expectation value above the ground state. The bounds lead to the Mandelstam-Tamm inequality and a quantum speed limit which is only slightly weaker than the Margolus-Levitin inequality.
\end{abstract}

\maketitle

\section{Introduction}\noindent
The recent paper by Jones and Kok \cite{jones10} presented alternative geometric derivations of the Mandelstam-Tamm \cite{mandelstam45} and Margolus-Levitin \cite{margolus98} inequalities for the quantum speed of dynamical evolution. The derivations were based on two independent bounds on the rate of change of the statistical distance between two arbitrary pure quantum states. Whereas the derivation of the Mandelstam-Tamm inequality is correct, the derivation of the Margolus-Levitin inequality (specifically, the derivation of a bound on the rate of change of the statistical distance) is wrong.
 
This Comment is organized as follows. In Sec.~\ref{sec::errors} we point out the error in the derivation of Eq.~(35) in Ref.~\cite{jones10}. In Secs.~\ref{sec::standard} and \ref{sec::mean}, we present two upper bounds on the rate of change of the statistical distance, expressed in terms of the standard deviation of the generator $K$ and its expectation value above the ground state, respectively. Finally, in Sec.~\ref{sec::mean} we also identify an immediate consequence of the new bound, namely, a quantum speed limit.

\section{Incorrect derivation}\label{sec::errors}\noindent
In the paper, the authors considered the unitary evolution parametrized by $\theta$ and generated by the Hermitian operator $K$. According to this evolution, a quantum system prepared in an initial pure state $|\psi_{0}\rangle$ evolves to
\be\label{finalstate}
|\psi_{\theta}\rangle = \exp\left(-\frac{i}{\hbar} K \theta\right) |\psi_{0}\rangle\, .
\ee
Let us point out a subtlety. The authors invoked the Wootters distance \cite{wootters81} between two pure states (which represents an angle between these states in Hilbert space), 
\be
s_{W}(\psi_{0},\psi_{\theta}) = \arccos(|\langle \psi_{0}|\psi_{\theta}\rangle|), \ \mbox{with} \  s_{W} \in \left[0, \frac{\pi}{2}\right]\, ,
\ee
and expressed the rate of change of the \textit{statistical} distance as a derivative of the Wootters distance with respect to $\theta$. However, the statistical distance between two pure states is defined as twice the Wootters distance \cite{anandan90}, that is,
\be\label{distance}
s = 2 s_{W}(\psi_{0},\psi_{\theta}) = 2 \arccos(|\langle \psi_{0}|\psi_{\theta}\rangle|), \ \mbox{with} \  s \in \left[0, \pi\right]\, .
\ee
Therefore, the rate of change of the statistical distance expressed in terms of the Wootters distance can be written as
\ba
\frac{ds}{d\theta} &=& 2\frac{d}{d\theta} \arccos(|\langle \psi_{0}|\psi_{\theta}\rangle|) \label{derivative} \\
&=& -\frac{2}{\sqrt{1-|\langle \psi_{0}|\psi_{\theta}\rangle|^2}} \frac{d}{d\theta} |\langle \psi_{0}|\psi_{\theta}\rangle|\, . \label{derivative2}
\ea
The missing factor of 2 is a minor point. Unfortunately then the authors transformed Eq.~(27) [our Eq.~(\ref{derivative2})] to Eq.~(28) by assuming that the derivative of $|\langle \psi_{0}|\psi_{\theta}\rangle|$ over the parameter $\theta$ is \textit{always} positive (and using $1/\sqrt{1-|\langle \psi_{0}|\psi_{\theta}\rangle|^2} \geq 1$). However, $|\langle \psi_{0}|\psi_{\theta}\rangle|$ can be a decreasing or increasing function of $\theta$ \cite{brody}. Furthermore, for small values of the parameter, $|\langle \psi_{0}|\psi_{\theta}\rangle|$ must decrease with $\theta$ (since $|\langle \psi_{0}|\psi_{\theta}\rangle|$ equals unity for $\theta=0$), and therefore its derivative is \textit{negative}. This invalidates the chain of relations following Eq.~(27) that led to the Margolus-Levitin inequality given in Eq.~(38). As a consequence, the extension of the derivation of the Margolus-Levitin inequality to the case of unitary evolutions of arbitrary mixed states presented in the paper does not hold either.

\section{Bound in terms of the standard deviation of the generator}\label{sec::standard}\noindent
In the paper, the authors introduced an upper bound on the rate of change of the statistical distance expressed in terms of the standard deviation of the generator $K$ by invoking the concept of Fisher information (see Eqs.~(22) and (23) in Ref.~\cite{jones10}). It is important to note at this point that in the paper two different statistical distances were considered (these distances are equal only in a special case discussed below). The (infinitesimal) statistical distance defined in Eq.~(14) [or equivalently in Eq.~(23)] in Ref.~\cite{jones10} is the (infinitesimal) distance along a given path generated by the Hermitian operator $K$ between the initial and final quantum states in the projective Hilbert space. This distance is measured by the Fubini-Study metric which is defined naturally from the inner product in Hilbert space \cite{anandan90} and furthermore can be related to the Fisher information. Whereas the statistical distance defined in Eq.~(15) in Ref.~\cite{jones10} [our Eq.~(\ref{distance}) with a missing factor of 2 included] is the distance along the shortest geodesic joining the initial and final quantum states in the projective Hilbert space \cite{anandan90}. In general, (the finite value of) the former distance is always greater than or equal to the latter distance \cite{anandan90} (the same relation applies to the absolute values of the rates of change of the respective statistical distances).

For the sake of completeness, we present a derivation of a bound \textit{analogous} to the bound given in Eq.~(23) in Ref.~\cite{jones10}; however, here we express the rate of change of the statistical distance as a derivative of the Wootters distance with respect to $\theta$. This bound results in the Mandelstam-Tamm inequality on the quantum speed of dynamical evolution. We begin by pointing out that in a paper by Bhattacharyya \cite{bhattacharyya83} the rate of change of the Wootters distance was shown to be upper-bounded by
\be
\frac{d}{d\theta} \arccos(|\langle \psi_{0}|\psi_{\theta}\rangle|) \leq \frac{\Delta K}{\hbar}\, ,
\ee
where $\Delta K$ is the standard deviation of $K$. Combining this bound with Eq.~(\ref{derivative}) yields
\be
\frac{ds}{d\theta} \leq \frac{2 \Delta K}{\hbar}\, , \label{bound}
\ee
which is \textit{analogous} to Eq.~(23) of Ref.~\cite{jones10} and (in the case where $\theta$ is the time parameter generated by the Hamiltonian $K \equiv H$) leads to the Mandelstam-Tamm inequality for the minimum time needed for a quantum system to evolve to an orthogonal state \cite{jones10}:
\be
t_{MT} \geq \frac{\pi}{2} \frac{\hbar}{\Delta E}\, . \label{MT}
\ee
The above bound on the rate of change of the statistical distance and the Mandelstam-Tamm inequality are saturated by the following \textit{optimal} states,
\be
|\psi\rangle = \frac{1}{\sqrt{2}} \left( |k_{\rm min}\rangle + e^{i \varphi} |k_{\rm max}\rangle \right)\, , \label{optimal}
\ee
where $|k_{\rm min}\rangle$ and $|k_{\rm max}\rangle$ are the eigenvectors corresponding to the maximal and minimal eigenvalues of the generator $K$ (here, we assume that the spectrum of the generator $K$ is upper- and lower-bounded). In the case of optimal states the two statistical distances discussed above are equal, which implies that the quantum system moves along a geodesic joining the initial and final states in the projective Hilbert space \cite{anandan90}.

\section{Bound in terms of the expectation value of the generator}\label{sec::mean}\noindent
Since the aim of the derivation in Ref.~\cite{jones10} was to obtain a bound on the rate of change of the statistical distance expressed in terms of the expectation value of the generator $K$ above the ground state, we show here how such a bound may indeed be obtained. In this Comment, we deal with upper bounds; therefore, it is not surprising that the rate of change of the statistical distance can be bounded in more than one way.

Let us begin with an observation. According to the unitary evolution governed by the generator $K$, a quantum system prepared in an initial pure state $|\psi_{0}\rangle$ evolves to the final pure state $|\psi_{\theta}\rangle$ as given in Eq.~(\ref{finalstate}). However, this expression does not take into account the freedom we have to multiply $|\psi_{\theta}\rangle$ by an arbitrary phase factor \cite{braunstein96},
\be
|\psi'_{\theta}\rangle =  e^{i f(K,\theta)} |\psi_{\theta}\rangle\, , \nonumber
\ee 
where $f(K, \theta) = h(K) + g(\theta)$ is a real-valued function with $h(K)$ and $g(\theta)$ denoting functions of the generator $K$ and parameter $\theta$, respectively. This form of $f(K, \theta)$ ensures that the overlap between the arbitrarily phase-shifted states is left unchanged, that is, $|\langle \psi'_{0}|\psi'_{\theta}\rangle| = |\langle \psi_{0}|\psi_{\theta}\rangle|$. The phase freedom in $|\psi_{\theta}\rangle$ corresponds to the freedom to rephase independently each of the eigenstates (eigenvalues) of $K$ without changing the statistical distance \cite{braunstein96}. A convenient phase choice,
\be
f(K, \theta) = g(\theta) =  \frac{K_{\rm min} \theta}{\hbar}\, ,
\ee
where $K_{\rm min}$ is the minimal eigenvalue of $K$, yields
\ba
|\psi'_{0}\rangle &=& |\psi_{0}\rangle\, , \nonumber \\
|\psi'_{\theta}\rangle &=& \exp\left( -\frac{i}{\hbar} (K - K_{\rm min}) \theta \right) |\psi_{0}\rangle \nonumber \, .
\ea 
Keeping in mind this subtlety, we begin the derivation by writing 
\ba
\frac{ds}{d\theta} &=& 2\frac{d}{d\theta} \arccos(|\langle \psi'_{0}|\psi'_{\theta}\rangle|)\nonumber \\
&=& -\frac{2}{\sqrt{1-|\langle \psi'_{0}|\psi'_{\theta}\rangle|^2}} \frac{d}{d\theta} |\langle \psi'_{0}|\psi'_{\theta}\rangle|\, .
\ea
From here we proceed along the similar lines as laid out in Ref.~\cite{jones10}; however, we retain the troublesome prefactor $1/\sqrt{1-|\langle \psi'_{0}|\psi'_{\theta}\rangle|^2}$ and pay special attention to the derivative of $|\langle \psi'_{0}|\psi'_{\theta}\rangle|$ over $\theta$. Thus, we write
\be\label{derivativeprime}
\frac{ds}{d\theta} \leq \left| \frac{ds}{d\theta} \right| = \frac{2}{\sqrt{1-|\langle \psi'_{0}|\psi'_{\theta}\rangle|^2}} \left| \frac{d}{d\theta} |\langle \psi'_{0}|\psi'_{\theta}\rangle | \right|\, .
\ee
Using the generalized Schr\"{o}dinger equation 
\be
i\hbar \frac{d}{d\theta} |\psi'_{\theta}\rangle = (K - K_{\rm min}) |\psi'_{\theta}\rangle\, ,
\ee
we can bound the derivative as
\ba
&&\left| \frac{d}{d\theta} |\langle \psi'_{0}|\psi'_{\theta}\rangle| \right| = \nonumber \\
&&= \frac{|i \langle \psi'_{0}|\psi'_{\theta}\rangle \langle \psi'_{\theta}|K - K_{\rm min} |\psi'_{0}\rangle - i \langle \psi'_{0}|K - K_{\rm min} |\psi'_{\theta}\rangle \langle \psi'_{\theta}|\psi'_{0}\rangle |}{2 \hbar |\langle \psi'_{0}|\psi'_{\theta}\rangle|} \nonumber \\ 
&&= \frac{| \mbox{Im} (\langle \psi'_{0}|K - K_{\rm min} |\psi'_{\theta}\rangle \langle \psi'_{\theta}|\psi'_{0}\rangle) |}{\hbar |\langle \psi'_{0}|\psi'_{\theta}\rangle|} \nonumber \\
&&\leq \frac{| \langle \psi'_{0}|K - K_{\rm min} |\psi'_{\theta}\rangle \langle \psi'_{\theta}|\psi'_{0}\rangle |}{\hbar |\langle \psi'_{0}|\psi'_{\theta}\rangle|} \nonumber \\
&&= \frac{| \langle \psi'_{0}|K - K_{\rm min} |\psi'_{\theta}\rangle | \cdot | \langle \psi'_{\theta}|\psi'_{0}\rangle |}{\hbar |\langle \psi'_{0}|\psi'_{\theta}\rangle|}\, , \nonumber
\ea
where in the last line we use $|ab| = |a| |b|$. We combine this bound with Eq.~(\ref{derivativeprime}) to obtain
\ba
\frac{ds}{d\theta} &\leq& \frac{2}{\sqrt{1-|\langle \psi'_{0}|\psi'_{\theta}\rangle|^2}} \frac{| \langle \psi'_{0}|K - K_{\rm min} |\psi'_{\theta}\rangle |}{\hbar} \nonumber \\
&\leq& \frac{2}{\sqrt{1-|\langle \psi'_{0}|\psi'_{\theta}\rangle|^2}} \frac{\langle \psi'_{0}|K - K_{\rm min}|\psi'_{0}\rangle}{\hbar} \nonumber \\
&=& \frac{2}{\sqrt{1-|\langle \psi'_{0}|\psi'_{\theta}\rangle|^2}} \frac{\langle K \rangle - K_{\rm min}}{\hbar}\, .
\ea
The second line can be verified directly by expanding $|\psi'_{0}\rangle$ and $|\psi'_{\theta}\rangle$ in the eigenbasis of $K$ and using $\cos x \leq 1$. Finally, we rewrite the denominator of the prefactor as
\be
\sqrt{1-|\langle \psi'_{0}|\psi'_{\theta}\rangle|^2} = \sin(s/2)\, , \label{fidelity}
\ee
where we use $|\langle \psi'_{0}|\psi'_{\theta}\rangle|^2 = |\langle \psi_{0}|\psi_{\theta}\rangle|^2 = \cos^{2}(s/2)$, with $s \in \left[0, \pi\right]$ [see also Eq.~(\ref{distance})] \cite{anandan90}. This yields
\be
\frac{ds}{d\theta} \leq \frac{2}{\sin(s/2)} \frac{\langle K \rangle - K_{\rm min}}{\hbar}\, . \label{newbound}
\ee
This is an upper bound on the rate of change of the statistical distance expressed in terms of the expectation value of the generator $K$ above the ground state (here, we assume that the spectrum of the generator $K$ is lower-bounded). A similar result for parameter (time)-dependent generators was derived by Deffner and Lutz \cite{deffner11}. 

Our result can be easily generalized to the following bound,
\be
\frac{ds}{d\theta} \leq \frac{2}{\sin(s/2)} \frac{\langle |K - \kappa| \rangle}{\hbar}\, , \label{newboundgeneral}
\ee
where $\kappa$ is some arbitrary real constant that appears in an appropriately defined phase factor: $f(K, \theta) = g(\theta) =  \kappa \theta/\hbar$. The bounds given in Eqs.~(\ref{newbound}) and (\ref{newboundgeneral}) apply to the unitary evolution of pure quantum states; however, these bounds can also be extended to the unitary evolution of arbitrary mixed states by employing a standard purification procedure \cite{jones10}.

Here, we use the bound given in Eq.~(\ref{newbound}) to derive a quantum speed limit. To this end, we separate the variables $s$ and $\theta$, and we integrate both sides:
\ba
\int^{\theta}_{0} d\theta' &\geq& \frac{\hbar}{2 (\langle K \rangle - K_{\rm min})} \int^{s_{\rm max}}_{0} \sin(s/2) \, ds \ \Rightarrow \nonumber \\
\theta &\geq& 2 \sin^{2}(s_{\rm max}/4) \, \frac{\hbar}{\langle K \rangle - K_{\rm min}}\, ,
\ea
where $s_{\rm max}$ is the maximal statistical distance traversed by the quantum system in Hilbert space. This is a generalized quantum speed limit. In the case where $\theta$ is the time parameter generated by the Hamiltonian $K \equiv H$ that evolves an initial state of the quantum system to its orthogonal counterpart, i.e., $s_{\rm max} = \pi$, we obtain the following quantum speed limit,
\be
t \geq \frac{\hbar}{E}\, ,
\ee
where $E = \langle H \rangle - H_{\rm min}$ is the average energy above the ground state of the quantum system. Note that this inequality is only slightly weaker than the Margolus-Levitin inequality \cite{margolus98}
\be
t_{ML} \geq \frac{\pi}{2} \frac{\hbar}{E}\, .
\ee
Similarly as in the case of the Mandelstam-Tamm inequality, the Margolus-Levitin inequality is saturated by the optimal states given in Eq.~(\ref{optimal}).

\section{Conclusions}
In summary, we provide a valid derivation of an upper bound on the rate of change of the statistical distance expressed in terms of the expectation value of the generator $K$ above the ground state. The bound results in a quantum speed limit that is only slightly weaker than the Margolus-Levitin inequality.

\begin{acknowledgements}\noindent
Valuable discussions with Pieter Kok, Michael Hall, Andy Chia, and Howard Wiseman are gratefully acknowledged. This research was supported by the ARC Centre of Excellence CE110001027.
\end{acknowledgements}

\end{document}